\documentclass[aps,
  prd,
  twocolumn,
  showpacs,
  nofootinbib,
  superscriptaddress
]{revtex4-1}

\usepackage{color}
\usepackage{graphicx}
\usepackage{latexsym}
\usepackage{float}
\usepackage{amsmath}
\usepackage{amssymb}
\usepackage{wasysym}
\usepackage{multirow}
\usepackage{epsfig}
\usepackage{hyperref}
\usepackage{subcaption}
\usepackage[export]{adjustbox}
\usepackage[utf8]{inputenc}

\newcommand{\be}{\begin{equation}}
\newcommand{\ee}{\end{equation}}
\newcommand{\ba}{\begin{eqnarray}}
\newcommand{\ea}{\end{eqnarray}}

\def\RGW{\rho_{\rm GW}}
\def\OGW{\Omega_{\rm GW}}
\def\Ophi{\Omega_{\rm \phi}}
\def\OSW{\Omega_{\rm SW}}
\def\OMHD{\Omega_{\rm turb}}

\def\d{{\rm d}}

\def\Tstar{T_\star}
\def\Hstar{H_\star}
\def\tstar{t_\star}

\def\RvacN{\rho_{\rm vac}}
\def\RradN{\rho_{\rm rad}(T_n)}

\def\fSW{f_{\text{SW}}}

\def\fSWs{f_{\text{SW},s}}
\def\fphis{f_{\phi,s}}
\def\fturbs{f_{\text{turb},s}}

\begin{document}

\title{Multi-wavelength observations of cosmological phase transitions using LISA and Cosmic Explorer}

\begin{abstract}
We reanalyze the detection possibilities for gravitational waves arising from cosmological first order phase transitions. We discuss the stochastic gravitational wave background corresponding to the three expected scenarios of phase transition dynamics. We then perform an analysis on the detection possibilities for each case using sensitivities for the next generation ground-based detector Cosmic Explorer and the current LISA proposal, using two analysis methods. We find that having both detectors allows wide detection possibilities over much of the parameter space, including those corresponding to several early Universe models.  
\end{abstract}

\author{Margot Fitz Axen$^a$, Sharan Banagiri$^a$, Andrew Matas$^a$, Chiara Caprini$^b$, Vuk Mandic$^a$}
\affiliation{$^a$School of Physics and Astronomy, University of Minnesota, Minneapolis, MN 55455, USA\\
$^b$Laboratoire Astroparticule et Cosmologie, CNRS UMR 7164, Universit$\acute{e}$ Paris-Diderot,  75013 Paris, France}

\maketitle

\section{Introduction}

First order cosmological phase transitions (PT) are predicted in scenarios beyond the standard model of particle physics, including in the context of the electroweak symmetry breaking (see e.g.~\cite{Kajantie:1996mn,Grojean:2006re,Caprini:2016re,Weir:2017wfa} and citations therein). Unlike second-order phase transitions in which the transition proceeds smoothly, first order phase transitions occur by the nucleation of bubbles of the new phase, which expand and collide. Energy released in the collisions along with bulk motion of any fluid present can give rise to a significant stochastic gravitational wave background \cite{Caprini:2016re,kosowsky1,kosowsky2,kosowsky3,kamionkowski,hindmarsh,caprinidurrer,caprinidurrer2,kahniashvili1,kahniashvili2,kahniashvili3}. For a review of cosmological sources of the stochastic gravitational-wave background, including PT models, see~\cite{Caprini:2018mtu}.

The background from the electroweak PT is a target for the Laser Interferometer Space Antenna (LISA) \cite{Grojean:2006re,LisaProposal:2017}. The most sensitive frequency band for LISA is $1-10\ {\rm mHz}$, and probes the Universe when it was at temperatures of $O(1\ {\rm TeV)}$, the expected scale of the electroweak PT \cite{Grojean:2006re,LisaProposal:2017}. A previous study in ~\cite{Caprini:2016re} has developed a phenomenological parameterization for the PT background, and estimated the range of models that can be detected by LISA, assuming four different LISA design proposals available at the time.

Unfortunately the electroweak PT background is not likely to be detectable by currently operating ground-based gravitational-wave detectors such as the Advanced Laser Interferometer Gravitational-wave Observatory (LIGO) and Advanced Virgo ~\cite{Stoch1:2015re,Stoch2:2014re,ALIGO:2015re}, which operate at roughly 10-5000 Hz and probe higher temperatures than LISA~\cite{Giblin:2014gra}. However, next generation ground-based detectors such as Cosmic Explorer (CE)\cite{Abbott:2017re} and the Einstein Telescope \cite{ETDesignDocument:2017} are expected to be roughly a factor of $10^3-10^4$ times more sensitive to the stochastic background, and will extend the observing band to lower frequencies $\sim$5 Hz \cite{Abbott:2017re,Sathyaprakash:2012jk}. Consequently, as we show below, CE may be able to measure the tail of the EW PT, or the signal from higher temperature PTs which could have spectra peaking near the CE band, if they exist \cite{Figueroa:2018xtu,Megias:2018sxv}.

In this study, we quantitatively assess what can be learned about the electroweak and other kinds of PTs by analyzing data from both LISA and CE simultaneously. A detection of the PT background by both LISA and CE would be a discovery of enormous significance, by providing a measurement of the energy density spectrum in two widely separated frequency bands. Even a null result in one detector could provide additional information about the PT background that the other detector could not achieve alone. 

We adopt the parameterization of PTs developed in~\cite{Caprini:2016re}, using the CE sensitivity from~\cite{Abbott:2017re} and LISA sensitivity from the most recent 2017 LISA proposal \cite{LisaProposal:2017}. The LISA noise estimates we use differ from those considered in~\cite{Caprini:2016re} in characteristics such as the arm length and mission duration. 

This study is organized as follows. In section 2, we discuss the projected sensitivities of Cosmic Explorer and LISA. In section 3 we provide a summary of the gravitational wave background produced by PTs. In section 4 we describe analysis methods used to determine the detectability of a gravitational wave (GW) signal. Finally, in section 5 we present the results of our analysis and discuss their implications. 

\section{Projected Sensitivity of LISA and Cosmic Explorer}

\begin{figure}[tp!]
\centering
\begin{subfigure}[t]{0.4\textwidth}
\includegraphics[width=\textwidth]{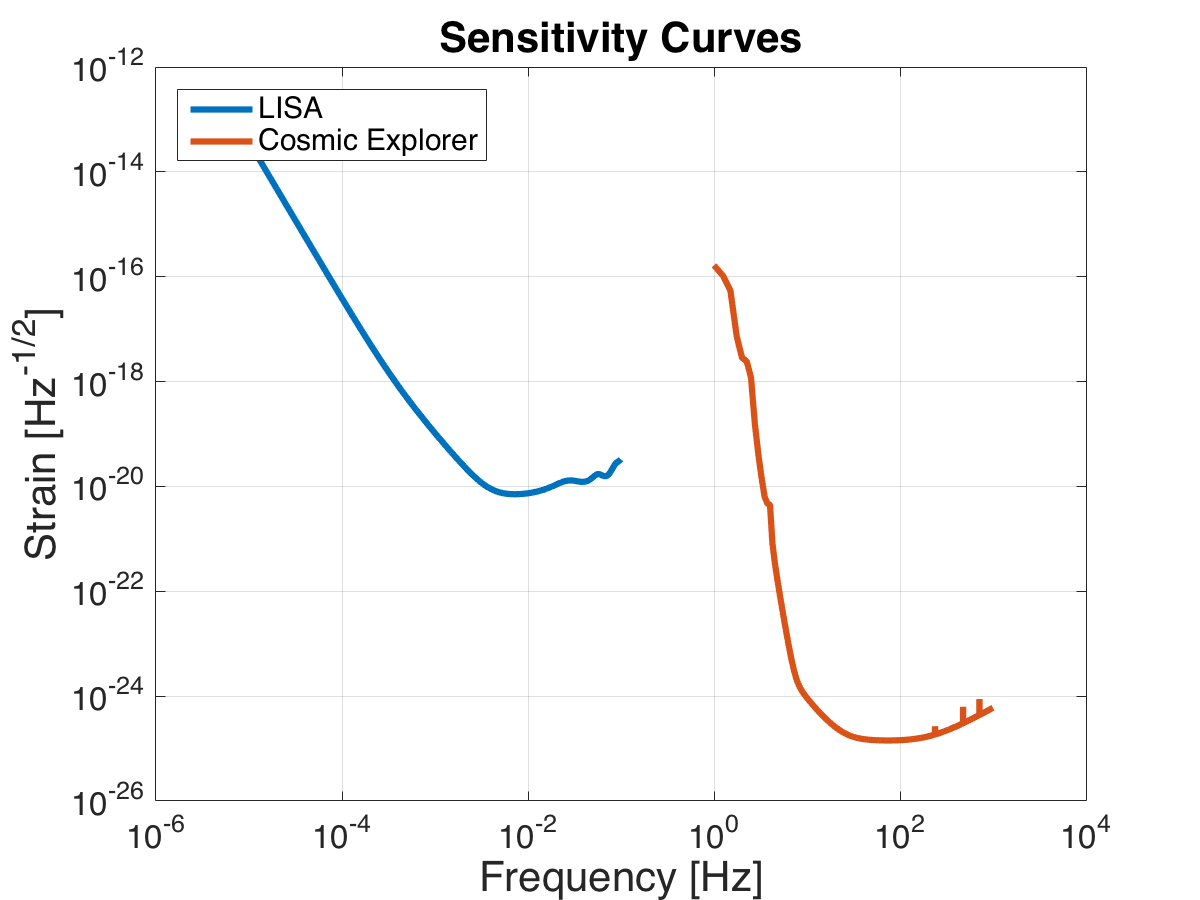}
\end{subfigure}
\caption{This plot shows the sky averaged sensitivity curves for LISA and CE, given by Equation 1.}
\label{ASD-curve}
\end{figure}

The 2017 LISA proposal~\cite{LisaProposal:2017} calls for a space-borne gravitational wave detector in a heliocentric orbit, lagging behind the Earth by 50-65 million km. The three LISA space-crafts will be in a triangular formation, with a separation of 2.5 million km between them. From the six interferometer links along LISA arms, three time delay interferometry (TDI) data channels (A, E, and T) will be constructed in order to perform cancellation of laser phase noise. To estimate the sensitivity of these channels, we use the acceleration and displacement noise specifications from the LISA proposal~\cite{LisaProposal:2017} and follow~\cite{Adams:2010re} to compute the minimum sky-averaged strain amplitude needed for a narrow-band signal at a frequency $f$ to stand above the noise. For a channel $I$ \footnote{In this paper we do not consider cross power spectra, since their sensitivities are generally much smaller than the auto-power spectra.} we equate the noise power spectral density $S^n_{I}(f)$ with the power induced by the gravitational wave, 
\begin{equation}
\mathcal{H} (f) \, \mathcal{R}_{I}(f) = S_{I}^n(f).
\end{equation}
Here $\mathcal{H} (f)$ is the power spectrum of the gravitational wave and $\mathcal{R}_{I}(f)$ is the sky-averaged detector response to the gravitational wave for channel $I$, which is related to the antenna function $F^A_I(\hat{\Omega},f)$ for the channel as, 
\begin{equation}
\mathcal{R}_{I}(f) = \sum_{A = +, \times} \int \frac{d \hat{\Omega}}{4 \pi} \, | F^A_{I} (\hat{\Omega}, f) |^2.
\end{equation}
Here $\hat\Omega$ depicts the solid angle. We note that the detector response captures effects due to the geometry of the detector, as well as the frequency dependence arising from the relative size of the detector in comparison with the wavelength of the gravitational wave. The strain sensitivity $h_I(f)$ is then just the square root of $\mathcal{H}(f)$, 
\begin{equation}
h(f) = \sqrt{\frac{S^n_{I} (f)}{\mathcal{R}_{I}(f)}}.
\label{eq:strain_sensitivity}
\end{equation}
We show the sky-averaged sensitivity curve for A and E channels (we assume $R_{A} = R_{E}$)
given by Equation~\ref{eq:strain_sensitivity} for LISA in Figure~\ref{ASD-curve}.  For more details about constructing LISA sensitivity curves, we refer the reader to~\cite{Cornish:2018dyw}.

Cosmic Explorer~\cite{Abbott:2017re} is a proposed 40 km long detector with a similar design to that of LIGO. It will extend the frequency band of Advanced LIGO and Advanced Virgo down to around 5 Hz, with sensitivity improvements of order $10\times$ relative to that of Advanced LIGO. Sensitivity curves for CE are computed from analytical models of noise sources including quantum noise, local gravitational disturbances, and thermal noise in suspensions and mirror coatings~\cite{Abbott:2017re}. Unlike for LISA, the CE detector response can be assumed to be independent of frequency because the wavelengths of the gravitational waves targeted by these detectors are much larger than the arm-length of the detector. Figure~\ref{ASD-curve} shows the CE target sensitivity.

\section{Phase Transition Stochastic Gravitational Wave Background}
The stochastic background is characterized by its spectral energy density
\begin{equation}
h^2\OGW(f) = h^2\frac{f}{\rho_c} \frac{\d\RGW}{\d f},
\end{equation}
where $\RGW$ is the energy density in gravitational waves, $\rho_c = 3H_0^2/(8\pi G)$ is the critical energy density to have a flat Universe, and the Hubble constant is $H_0=100 h\ {\rm km/s/Mpc}$ (we adopt the speed of light $c=1$).
The source of gravitational wave energy arising from first-order cosmological PTs is due to bubble collisions and fluid motion. In Ref.~\cite{Caprini:2016re}, three different processes are considered, whose different contributions to the total background must be calculated separately and summed. However, depending on the dynamics of the phase transition, some of these processes may provide a negligible contribution to the entire background and so can be disregarded.   

The first contribution is that from the scalar field, $\phi$, itself, due to the bubble wall collisions.
The second contribution is from sound waves in the plasma, as the bubble wall sweeps through the surrounding fluid. 
The final contribution is from magneto-hydrodynamic (MHD) turbulence in the plasma. 
We denote these three contributions $\Ophi, \OSW$, and $\OMHD$, respectively. In general, the total background is~\cite{Caprini:2016re} 
\begin{equation}
h^2\OGW(f) = h^2\Ophi(f) + h^2\OSW(f) + h^2\OMHD(f).
\end{equation}
The relative importance of these three terms depends on the dynamics of the PT.

Following \cite{Caprini:2016re}, in this study we consider three different scenarios for the dynamics of the bubble expansion. The first scenario assumes non-runaway bubbles whose speed reaches a relativistic terminal velocity. The second scenario assumes runaway bubbles which, though expanding in the plasma, rapidly approach the speed of light. The third scenario assumes that the phase transition occurs in vacuum, consequently plasma effects are negligible and the bubbles expand at the speed of light. Note that the second scenario is excluded in the context of the electroweak symmetry breaking \cite{Bodeker:2017re}: we therefore apply it here only for the case of speculative PTs occurring at very high temperatures, relevant for observations with the CE. 

\subsection{Main parameters describing the phase transition}
We first briefly review the parameterization used throughout the rest of the study, following~\cite{Caprini:2016re}, and the relevant quantities for computing the gravitational wave background. As usual we define $\beta$ as the inverse of the time duration of the phase transition. This quantity determines the size of the bubbles at the time of collision and therefore the characteristic frequency at which the GW signal peaks~\cite{Grojean:2006re}. For a phase transition taking place at temperature $T_{n}$ and ending at time $t_n$ \cite{Grojean:2006re}, $\beta$ is given by \cite{Caprini:2016re} 
\begin{equation}
\beta = \frac{\Gamma'}{\Gamma}=-\left.\frac{\d S}{\d t}\right|_{t=t_n}=T_{n}H(T_{n})\left.\frac{\d S}{\d T}\right|_{T=T_n}
\end{equation}
where $\Gamma(t)$ denotes the nucleation rate and S is the Euclidean action of a bubble \cite{Grojean:2006re}. The last expression is given by the fact that $\d T/\d t = -TH$, where $H$ is the Hubble parameter \cite{Grojean:2006re}.

We define $\Tstar$ as the temperature of the thermal bath at the time $\tstar$ when gravitational waves are produced. A key parameter controlling the gravitational wave spectrum is $\beta/\Hstar$, where $\Hstar$ is the Hubble parameter at $\Tstar$. 

Another key parameter is $\alpha$, the ratio of the latent heat released during the phase transition to that of the radiation bath. It is given by \cite{Grojean:2006re}
\begin{equation}
\alpha=\frac{\RvacN}{\RradN},
\end{equation}
where $\RradN= g_n\pi^2T_n^4/30$ and $g_n$ is the number of relativistic degrees of freedom in the plasma at temperature $T_n$ \cite{Caprini:2016re}.

\subsection{Case 1: Non Runaway Bubbles}

The first phase transition scenario considers non-runaway bubbles, which expand in the plasma and reach a terminal velocity $v_w$ that is less than the speed of light. In this case, there are no large reheating effects and so $T_n \approx \Tstar$ \cite{Caprini:2016re}. There are two contributions to the gravitational wave spectrum that should be considered, due to sound waves and magnetohydrodynamic turbulence in the plasma after the bubbles have collided \cite{Caprini:2016re}
\begin{equation} h^2\Omega_{\text{GW}}(f) \approx h^2\Omega_{\text{SW}}(f)+h^2\Omega_{\text{turb}}(f).
\end{equation}
Gravitational waves from the scalar field play a negligible contribution in this case 
\cite{hindmarsh,hindmarsh2,hindmarsh3}.

Sound waves are generated by the bubble growth, and propagate through the plasma after the transition has completed \cite{hindmarsh,hindmarsh2,hindmarsh3}. A model covering all relevant values of $v_w$ and $\alpha$ is unavailable; however, simulations in \cite{hindmarsh2,hindmarsh3} give insights into the possible frequency dependence of the sound wave GW spectrum. We follow \cite{Caprini:2016re} which adopts the spectral shape
\begin{equation}
S_{\text{SW}}(f)=\left(\frac{f}{\fSW}\right)^3\left(\frac{7}{4+3(f/\fSW)^2}\right)^{7/2},
\end{equation}
where the observed frequency $f$ is related to the source frequency $f_s$ by $f=f_s/(1+z)$.
The overall scale of the sound wave peak frequency $\fSW$ is $\fSWs=1.15\beta/v_w$, a conservative estimate that agrees with the above spectral shape \cite{Caprini:2016re}.
The peak frequency $\fSW$ of the observed gravitational wave spectrum is given by
\begin{equation}\label{eq:fred}
\fSW=\frac{\fSWs}{1+z} = h_\star\left(\frac{\fSWs}{\beta}\right)\left(\frac{\beta}{\Hstar}\right),
\end{equation}
where $h_\star$ is the value of the inverse Hubble time at GW production redshifted to today 
\begin{equation}
h_\star=\frac{H_\star}{1+z} = 16.5 \times10^{-3}\text{mHz}\left(\frac{\Tstar}{100 \text{ GeV}}\right)\left(\frac{g_\star}{100}\right)^{\frac{1}{6}}.
\end{equation}
Finally, results from \cite{hindmarsh2} are fitted reasonably by the following gravitational wave spectrum
\begin{eqnarray} 
h^2\Omega_{\text{SW}}(f) = 2.65 \times 10^{-6} \left(\frac{\Hstar}{\beta}\right) \left(\frac{\kappa_{\nu}\alpha}{1+\alpha}\right)^2 \nonumber
\\
\times \left(\frac{100}{g_\star}\right)^{\frac{1}{3}}v_wS_{\text{SW}}(f),
\end{eqnarray}
where $\kappa_{\nu} = \rho_{\nu}/\rho_{\text{vac}}$ is the fraction of vacuum energy that gets converted into bulk motion of the fluid.
In the limits of large $v_w$, $\kappa_{\nu}$ is approximately given by \cite{Espinosa:2010re}
\begin{equation} 
\kappa_{\nu}\approx \alpha(0.73+0.083\sqrt{\alpha}+\alpha)^{-1}
\label{eq:kappa}
\end{equation}

In addition to sound waves, bubble percolation can also cause turbulence in the plasma, and in particular MHD turbulence since the plasma is ionized. For the GW signal from MHD turbulence, we adopt the spectral shape found analytically in \cite{Caprini:2009re} given by \cite{Caprini:2016re}
\begin{equation} S_{\text{turb}}(f)=\frac{(f/f_{\text{turb}})^3}{[1+(f/f_{\text{turb}})]^{11/3}(1+8\pi f/h_\star)} \end{equation}
Like the sound wave case, the peak frequency for the gravitational wave spectrum depends on the bubble size at the end of the transition and is given by 
$\fturbs=1.75\beta/v_w$ \cite{Caprini:2016re,Caprini:2009re}. Finally, the total contribution to the gravitational wave spectrum can be modelled as \cite{Caprini:2009re}
\begin{eqnarray}
h^2\Omega_{\text{turb}}(f) = 3.35 \times 10^{-4} \left(\frac{H_\star}{\beta}\right) \left(\frac{\kappa_{\text{turb}}\alpha}{1+\alpha}\right)^{\frac{3}{2}} \nonumber
\\
\times \left(\frac{100}{g_\star}\right)^{\frac{1}{3}}v_wS_{\text{turb}}(f), 
\end{eqnarray}
where the factor $\kappa_{\text{turb}}=\epsilon\kappa_{\nu}$ represents the fraction of the bulk motion that is turbulent.

\subsection{Case 2: Runaway Bubbles in Plasma}
The second case we consider is runaway bubbles in plasma, for which the bubble wall velocity $v_w$ approaches the speed of light. This scenario for the bubble expansion is not realised in the context of the electroweak PT \cite{Bodeker:2017re}, but it is in principle allowed in potential phase transitions occurring at higher temperature, which must be considered in the present analysis since they are relevant for the CE, as we will see. In this case, the contribution to the spectrum from the scalar field must be added to that from sound waves and turbulence \cite{Caprini:2016re}. 
\begin{equation}
h^2\OGW(f) \approx h^2\Ophi(f)+h^2\OSW+h^2\OMHD.
\end{equation}
Numerical simulations have been done to determine the contribution to the gravitational wave signal from the scalar field in \cite{Huber:2008re}, and the spectral shape of the gravitational wave spectrum is given by
\begin{equation} \label{Sphi}
S_{\phi}(f) = \frac{3.8(f/f_{\phi})^{2.8}}{1+2.8(f/f_{\phi})^{3.8}}. 
\end{equation}
The peak frequency from the scalar field is 
$\fphis=0.62\beta/(1.8-0.1v_w+v_w^2)$~\cite{Caprini:2016re,Huber:2008re}. Fits to simulation data give the total contribution to the gravitational wave spectrum as \cite{Caprini:2016re,Huber:2008re}
\begin{eqnarray}
\label{eq:Omphispec_case2} h^2\Omega_{\phi}(f) &=&1.67 \times 10^{-5} \left(\frac{H_\star}{\beta}\right)^2 \left(\frac{\kappa_{\phi}\alpha}{1+\alpha}\right)^2 \left(\frac{100}{g_\star}\right)^{\frac{1}{3}} \nonumber \\
&& \times \left(\frac{0.11v_w^3}{0.42+v_w^2}\right)S_{\phi}(f),
\end{eqnarray}
where the parameter $\kappa_{\phi}=\rho_{\phi}/\rho_{\text{vac}}$ is the fraction of vacuum energy that gets converted into energy of the scalar field. For this case, it is necessary to define a new parameter $\alpha_{\infty}$ as the minimum value of $\alpha$ such that bubbles run away \cite{Espinosa:2010re}.  
For $\alpha>\alpha_{\infty}$, 
the contribution of the scalar field to the gravitational wave background is parametrized  by \cite{Espinosa:2010re}
\begin{equation}
\kappa_\phi=1-\frac{\alpha_{\infty}}{\alpha}\geq 0
\end{equation}
In this expression, $\alpha_{\infty}/{\alpha}$ is the fraction of the total energy that goes into bulk motion ($\kappa_{\nu}$) and thermal energy ($\kappa_{\text{therm}}$); the amount of this energy that goes into bulk motion is given by \cite{Espinosa:2010re}
\begin{equation}
\kappa_\nu=\frac{\alpha_\infty}{\alpha}\kappa_{\infty}
\end{equation}
where $\kappa_{\infty}$ is computed similarly to Equation \ref{eq:kappa} as \cite{Espinosa:2010re}
\begin{equation} 
\kappa_{\infty}\approx \alpha_{\infty}(0.73+0.083\sqrt{\alpha_{\infty}}+\alpha_{\infty})^{-1}
\end{equation}

\subsection{Case 3: Runaway Bubbles in Vacuum}

The final case we consider is runaway bubbles in a vacuum dominated epoch, for which one  
only needs to consider the contribution to the spectrum from the scalar field and not from sound waves or turbulence, as those contributions are only applicable in plasma \cite{Caprini:2016re}
\begin{equation}
h^2\OGW(f) \approx h^2\Ophi(f).
\end{equation}
The spectral shape of the gravitational wave spectrum is given by Eq.~\ref{Sphi}. 
Furthermore, since $T_n$ goes to 0, the parameter $\alpha$ approaches infinity and therefore drops out of the expression for $\OGW(f)$. In this limit, the total contribution to the gravitational wave spectrum is therefore \cite{Caprini:2016re,Huber:2008re}
\begin{eqnarray}
\label{eq:Omphispec_case3} h^2\Omega_{\phi}(f) &=&1.67 \times 10^{-5} \left(\frac{H_\star}{\beta}\right)^2 \left(\frac{100}{g_\star}\right)^{\frac{1}{3}} \nonumber
\\
&& \times \left(\frac{0.11v_w^3}{0.42+v_w^2}\right)S_{\phi}(f).
\end{eqnarray}

\begin{figure*}[tp!]
\centering
\begin{subfigure}[t]{0.3\textwidth}
\includegraphics[width=\textwidth]{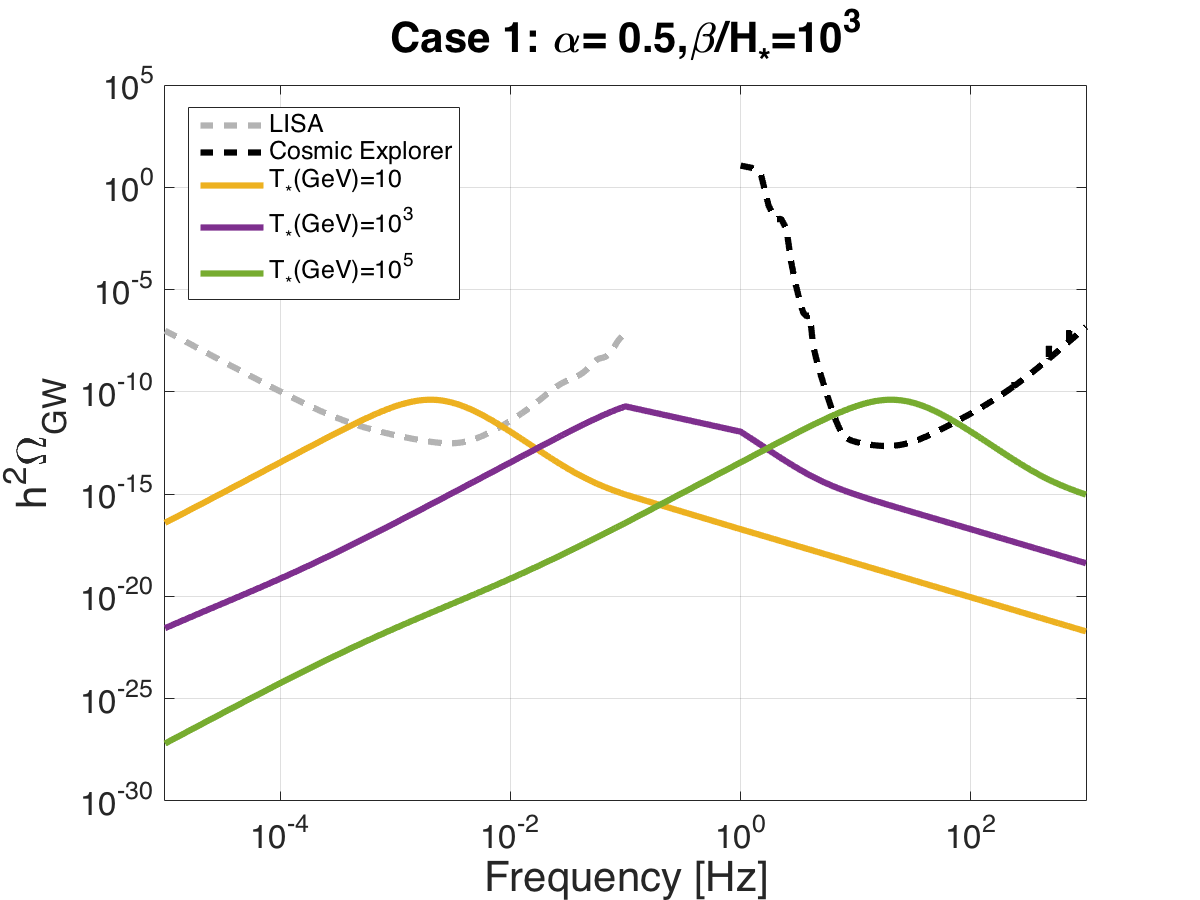}
\end{subfigure}
\begin{subfigure}[t]{0.3\textwidth}
\includegraphics[width=\textwidth]{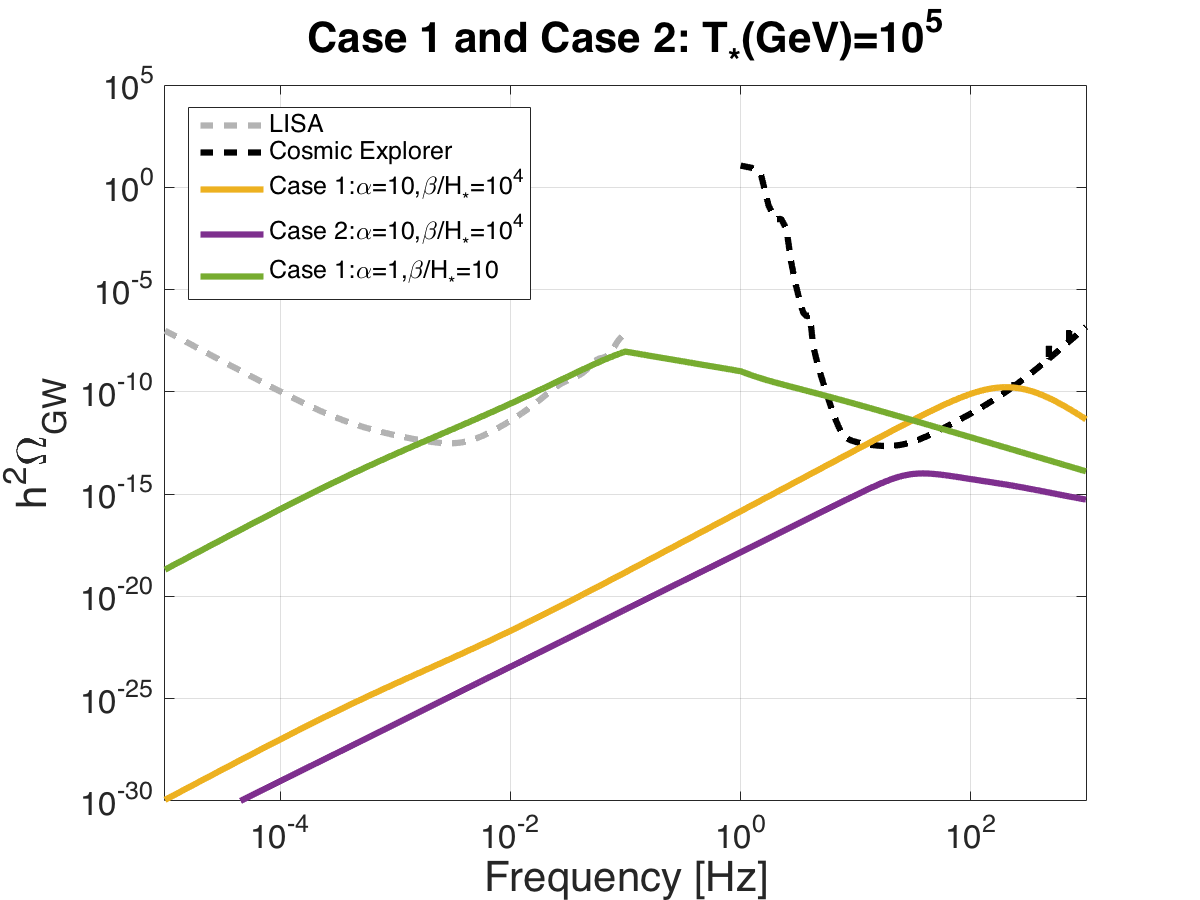}
\end{subfigure}
\begin{subfigure}[t]{0.3\textwidth}
\includegraphics[width=\textwidth]{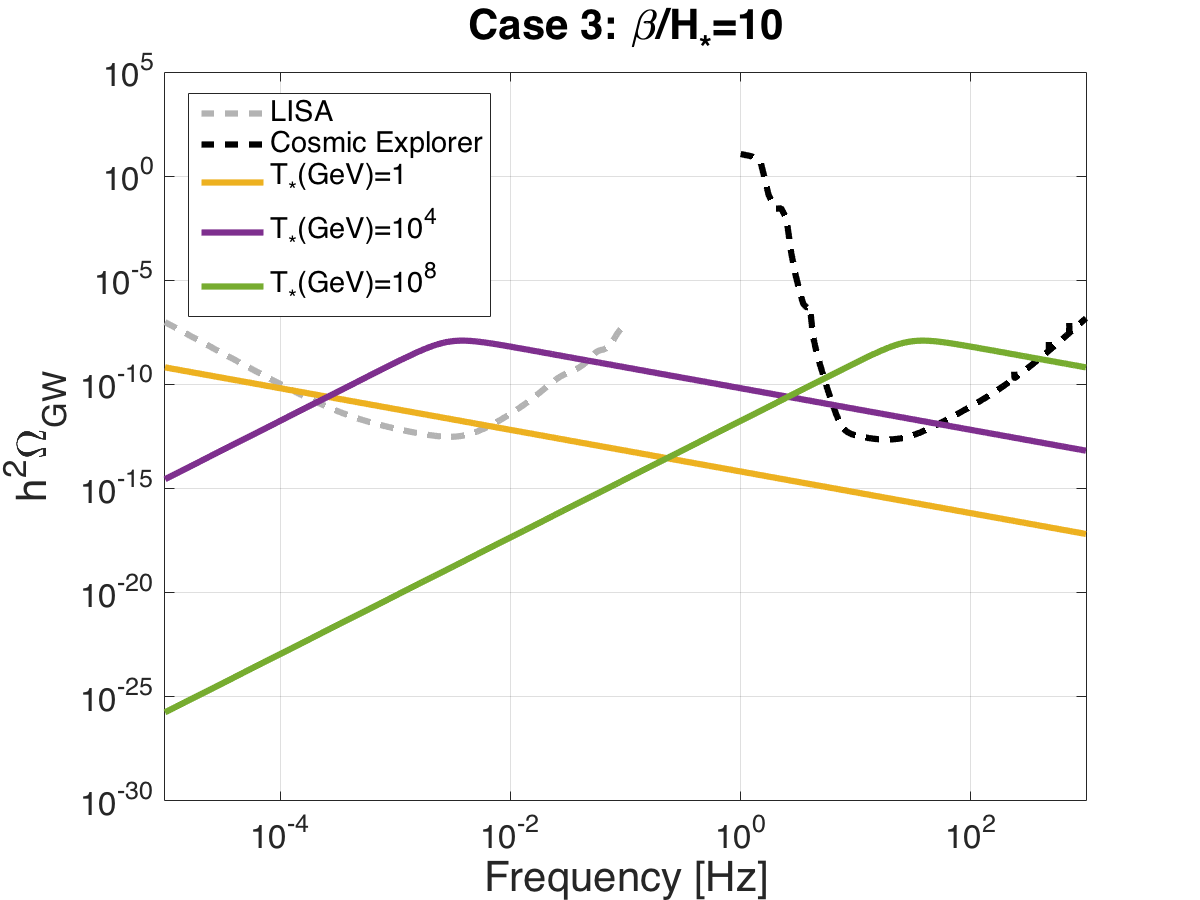}
\end{subfigure}
\caption{These plots show the total gravitational wave spectrum for specific points in the parameter space along with time integrated sensitivity curves for LISA and Cosmic Explorer (assuming 1 year of integration time). For the first and third plots, the point in the parameter space (respectively $(\alpha,\,\beta/H_\star)$ and $\beta/H_\star$ alone) corresponding to the parameter values indicated in the title of each plot, is computed for three different temperatures (corresponding to the yellow, purple, and green curves). The middle plot shows indicated points of interest in the parameter space of Cases 1 and 2 at $T_\star=100000$ GeV. These points are shown by the blue diamonds in Figures \ref{contours} and \ref{contours2}.}
\label{Omega-curve}
\end{figure*}

The computed gravitational-wave backgrounds for certain points in the parameter space of each phase transition case are shown in Figure~\ref{Omega-curve}, along with the sensitivity curves for LISA and Cosmic Explorer, assuming 1 year of exposure. For Cosmic Explorer, we assume two detectors are built at locations yielding the same overlap reduction function as for the two LIGO detectors. These figures provide an estimate of whether the selected models produce a spectrum that is large enough for detection in the frequency bands of LISA or Cosmic Explorer or both. In general, one expects that a spectrum rising well above a sensitivity curve should be detectable by the corresponding detector. Similarly, a spectrum well below a sensitivity curve is likely undetectable by that detector.  

\section{Analysis Methods}

In this section, we consider how to assess our ability to detect the gravitational wave background from first order PTs. We desire the analysis to be independent of any specific phase transition model, so relevant parameter values vary freely.

The three phase transition scenarios must be considered independently, as they have different gravitational wave spectra. For the case of runaway bubbles in vacuum (Case 3), there are only two parameters to be considered- $\beta/H_\star$ and $T_\star$. All other parameters are irrelevant at maximal bubble wall velocity and without plasma effects. 

For non-runaway bubbles in plasma (Case 1), however, there are five parameters that may be varied: $\beta/H_\star$, $T_\star$, $\alpha$, $v_w$, and $\epsilon$. The bubble wall velocity $v_w$ is model dependent but, as done in \cite{Caprini:2016re}, we chose to fix this parameter to be $v_w = 0.95c$ since lower wall velocities produce less observable gravitational radiation. The fraction of bulk motion that is turbulent, represented  by the parameter $\epsilon$, is also model dependent (and not yet fully understood). We set $\epsilon = 1$, corresponding to equal amount of kinetic energy in the sound waves and turbulent fluid motions: a plausible value for the case of strong enough PTs. The three remaining parameters, $\alpha$, $\beta/H_\star$, and $T_\star$ are related but again cannot be specified without choosing a phase transition model. As done in previous analyses these are the parameters we chose to vary: specifically, we analyze the $\alpha$-$\beta/H_\star$ space for several values of $T_\star$ \cite{Caprini:2016re,Grojean:2006re,binetruy}. 

For runaway bubbles in plasma (Case 2), the situation is similar. In this case, the five parameters to be considered are $\beta/H_\star$, $T_\star$, $\alpha$, $\alpha_{\infty}$, and $\epsilon$. The value of $\alpha_{\infty}$ is again model dependent. We choose to fix it to $\alpha_{\infty}=0.1$, a value considered also in \cite{Caprini:2016re}. In addition, in the context of case 2 we fix the temperature to $T_\star=10^5$ GeV, corresponding to an unknown PT at high temperature for which the presence of runaway bubbles in the plasma cannot be excluded. We then analyze the $\alpha$-$\beta/H_\star$ space for these parameter values just as for Case 1.

We scan the two-dimensional parameter space and compute the  total SGWB at 2500 points. Then at each point we use the projected sensitivities for LISA and Cosmic Explorer to evaluate whether the parameter values at the point describe a gravitational wave background that should be detectable by LISA or CE. We also determine the possibility of detection with the sensitivity of both detectors operating simultaneously. 

\subsection{Likelihood Analysis}

To analyze whether a point in the parameter space was in a detectable region, we considered two analysis methods. The first is a Bayesian likelihood analysis. 
We first define a projected sensitivity $\Omega_{\text{Sens}}$, which is related to the sky averaged sensitivity seen in  Equation~\ref{eq:strain_sensitivity} by \cite{Adams:2010re}
\begin{equation} 
\Omega_{\text{Sens}}=\frac{4\pi^2f^3}{3H_0^2}h^2_{I}(f).
\end{equation}
Inspired by similar analyses performed by LIGO~\cite{Mandic:2012pj}, we then define the following likelihood function for each point in the parameter space
\begin{equation}\label{eq:lik}\text{log}(L(\alpha, \beta/H_\star, \Tstar)) = -\tau\sum\limits_{f=f_{\text{min}}}^ {f_{\text{max}}}\left[\frac{h^2\Omega_{\text{GW}}( \alpha, \beta/H_\star,\Tstar,f)^2}{2h^2\Omega_{\text{Sens}}(f)^2}\right]\end{equation}
where $h^2\Omega_{\text{GW}}$ is the calculated GW signal and $\tau$ is the duration of the mission, assumed to be one year for this study. The summation over frequency runs over either the LISA frequency band, CE frequency band, or frequency band of both for evaluation of the combined sensitivity of both detectors operating simultaneously.

We assume uniform priors in input parameters $\alpha, \beta/H_\star, \Tstar$, implying that the Bayesian posterior distribution in these parameters is equal to the likelihood function defined above. We define a set of contours labeled by $Z$, for which the posterior is equal to $Z$. We then define the fraction of the total posterior probability contained within each contour,
\begin{equation} P(Z) = \frac{ \int\limits_{L(\alpha, \beta/H_\star, T_{\star})>Z} L(\alpha, \beta/H_\star, T_{\star}) \; d\alpha \; d(\beta/H_\star) \; dT_\star}{\int L(\alpha, \beta/H_\star,T_{\star}) \; d\alpha \; d(\beta/H_\star) \; dT_\star} \end{equation}
and identify the 95\% confidence contour as the one for which $P(Z)=0.95$.

\subsection{SNR Analysis}

In addition to the Bayesian likelihood analysis, we also considered a second method of analysis involving computing an SNR. Using the projected sensitivities of LISA and Cosmic Explorer, we consider the signal-to-noise ratio at each point of the parameter space \cite{Caprini:2016re}

\begin{equation} \label{eq:SNR}\text{SNR}(\alpha, \beta/H_\star, T_{\star})= \sqrt{\tau\int_{f_{\text{min}}}^{f_{\text{max}}} df \left[\frac{h^2 \Omega_{\text{GW}}( \alpha, \beta/H_\star,T_{\star},f)^2}{h^2\Omega_{\text{Sens}}(f)^2}\right]}\end{equation}

If the SNR value is greater than a threshold value $\text{SNR}_{\text{thr}}$, then the signal at that point is detectable. Contours are made outlining all points in this detectable region. 

Equations \ref{eq:lik} and \ref{eq:SNR} are in appearance very similar. In fact, the likelihood computation is analogous to doing the SNR calculation using a value of $\text{SNR}_{\text{thr}}=2$. However, due to the differences in the way contours are computed, 
and in particular the fact that the likelihood is defined over a uniform prior on the parameters $\{\alpha,\beta/H_\star,T_\star\}$ which are non-linearly related to $\OGW(f)$, they will produce slightly different results. In choosing a value of $\text{SNR}_{\text{thr}}$, we first follow \cite{Caprini:2016re} and consider a value of $\text{SNR}_{\text{thr}}=10$, and then compute contours for $\text{SNR}_{\text{thr}}=2$ for appropriate comparison to the likelihood analysis. Discrepancies between our calculations and those done in \cite{Caprini:2016re} for LISA are due to using a different value of $\epsilon$ along with using the most recent LISA sensitivity curves.

\section{Results and Discussion}

\begin{figure*}[tp!]
\centering
\begin{subfigure}[t]{0.4\textwidth}
\includegraphics[width=\textwidth]{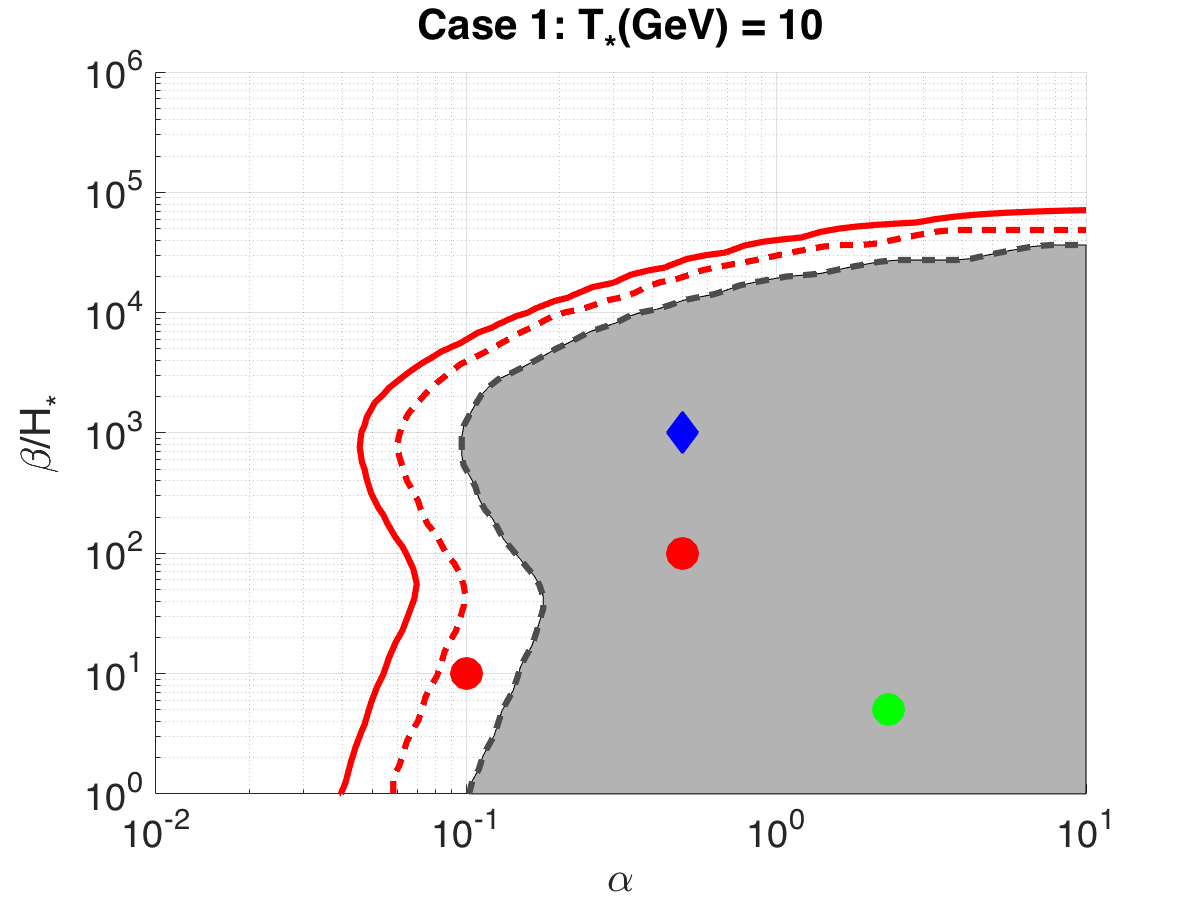}
\end{subfigure}
\begin{subfigure}[t]{0.4\textwidth}
\includegraphics[width=\textwidth]{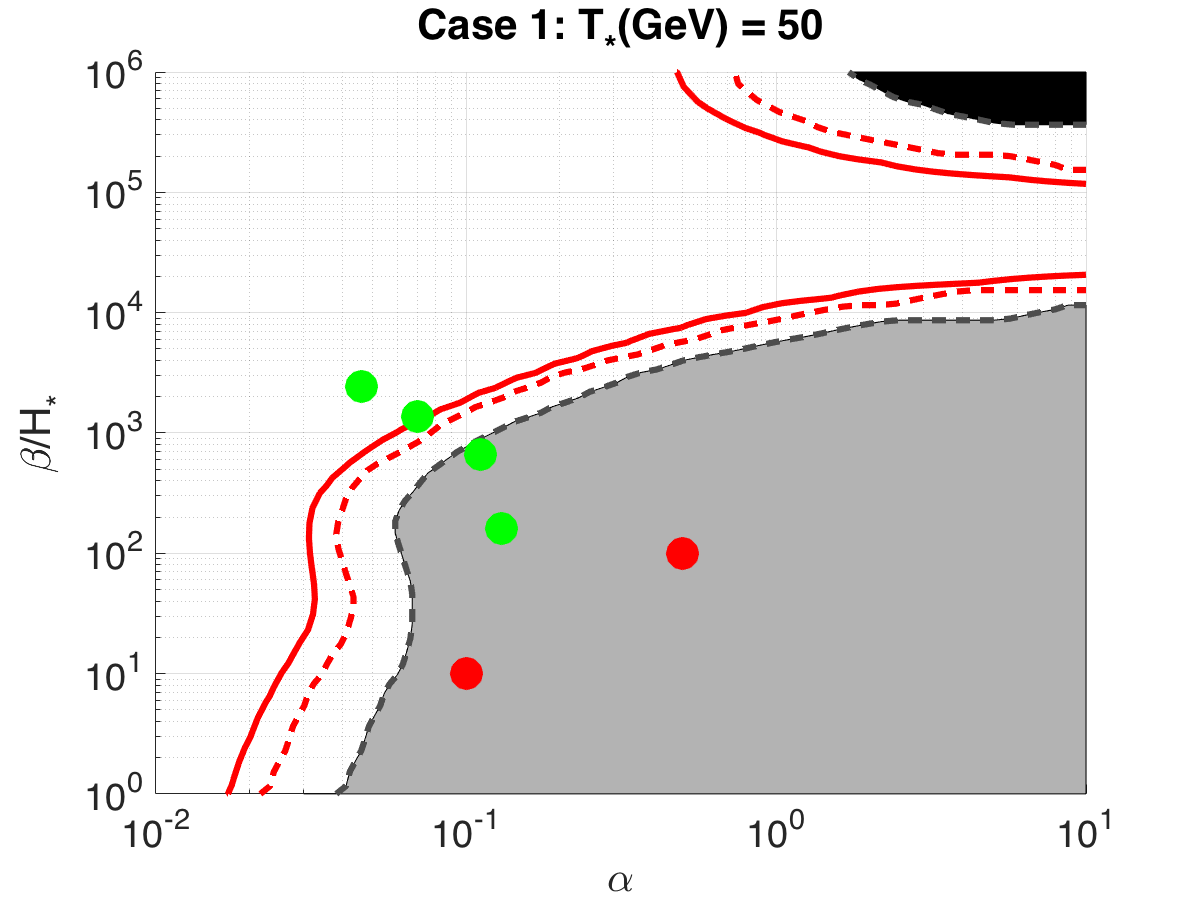}
\end{subfigure}
\begin{subfigure}[t]{0.4\textwidth}
\includegraphics[width=\textwidth]{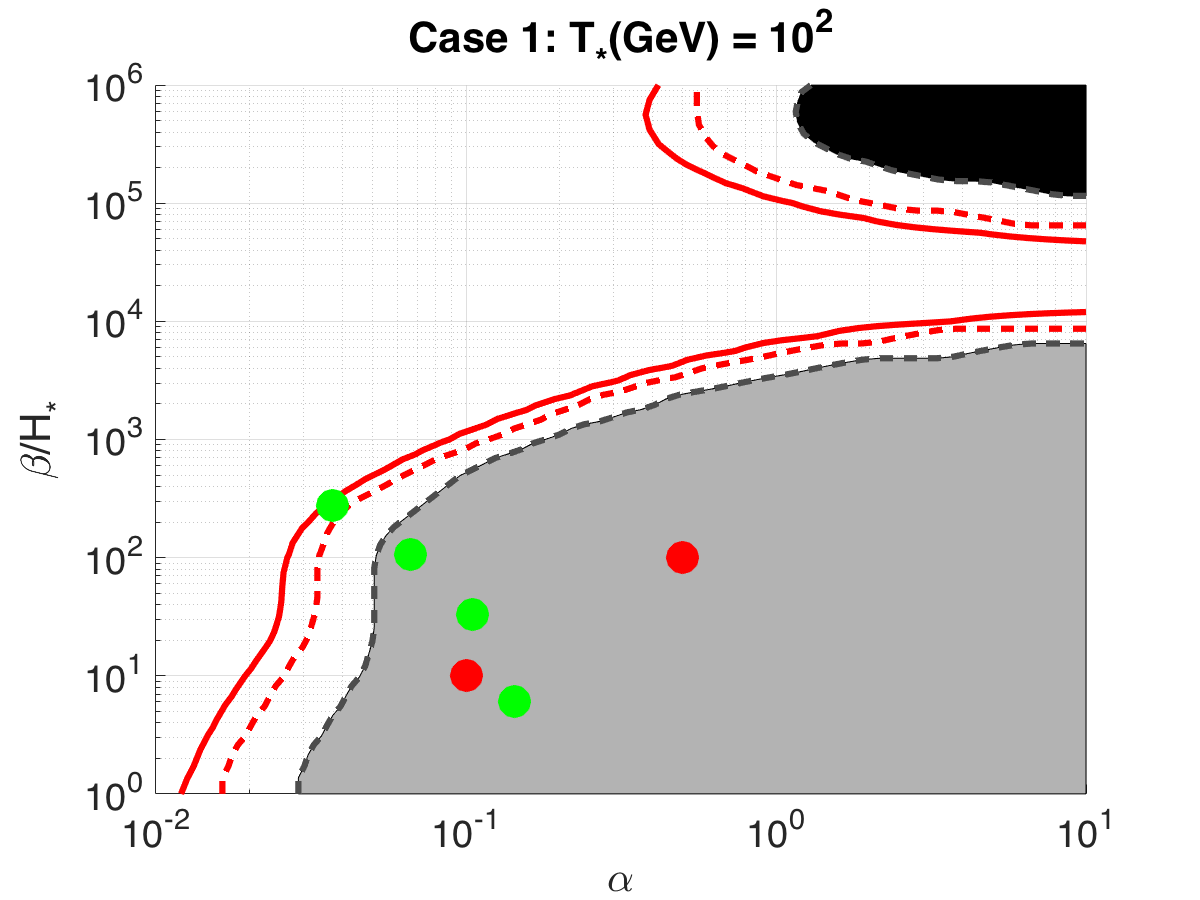}
\end{subfigure}
\begin{subfigure}[t]{0.4\textwidth}
\includegraphics[width=\textwidth]{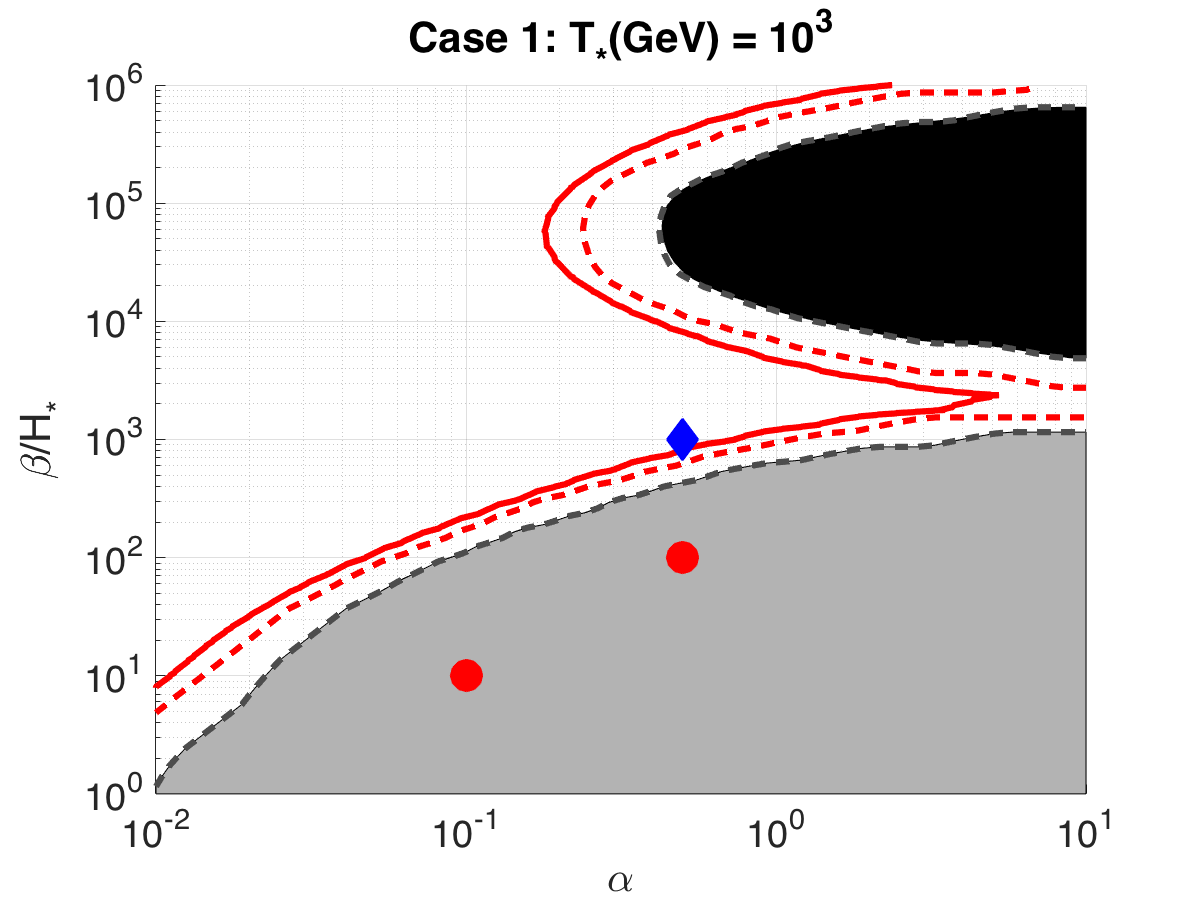}
\end{subfigure}
\begin{subfigure}[t]{0.4\textwidth}
\includegraphics[width=\textwidth]{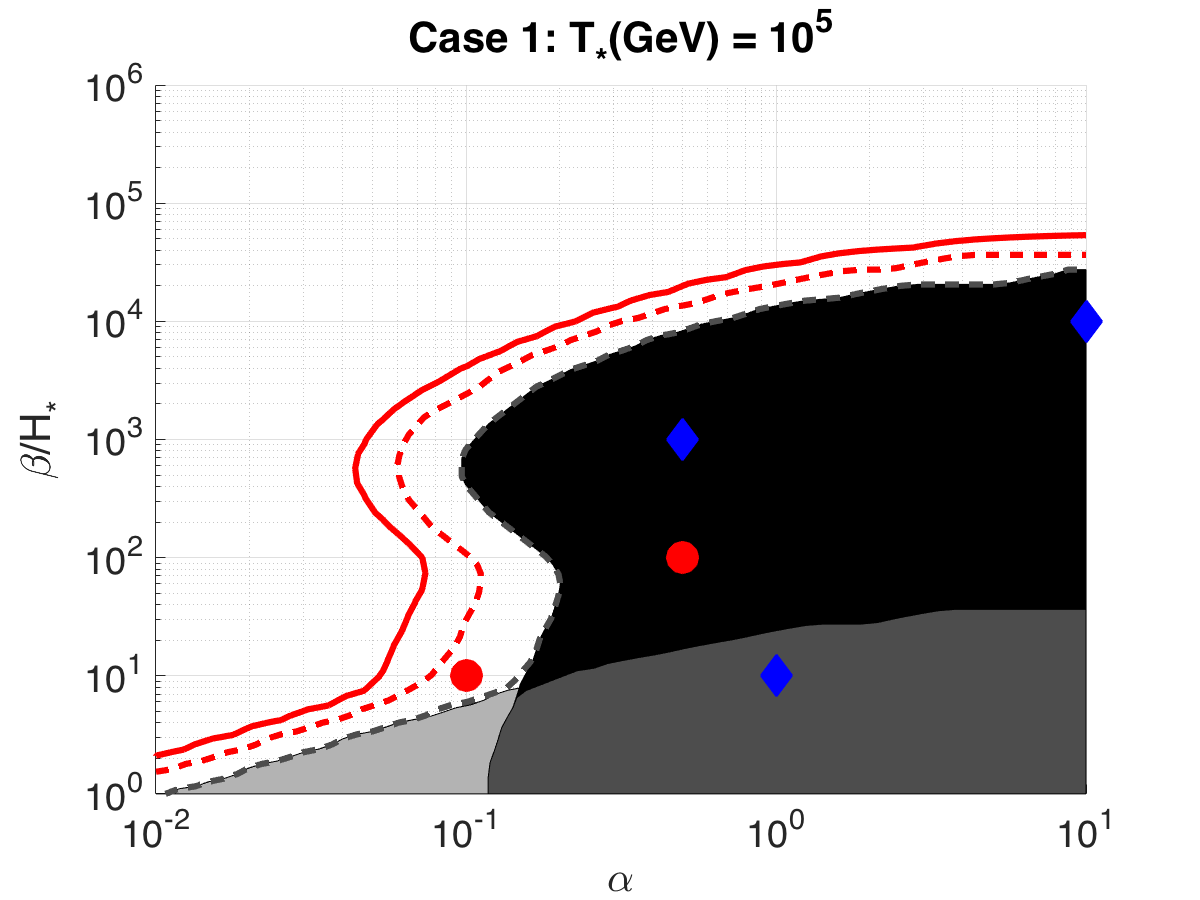}
\end{subfigure}
\begin{subfigure}[t]{0.35\textwidth}
\includegraphics[width=\textwidth]{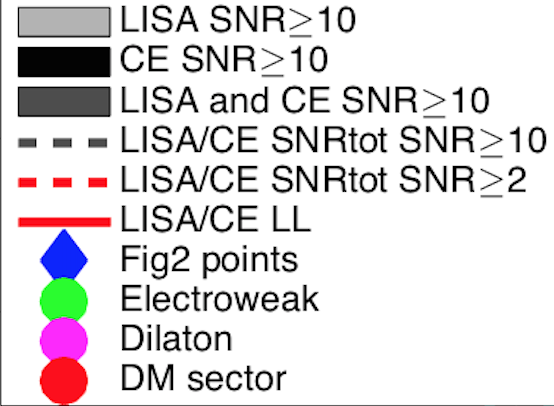}
\end{subfigure}
\caption{Shown here are five contour plots denoting different temperatures for Case 1. Shaded regions correspond to the total parameter space accessible to LISA, CE, or both, for the SNR analysis done with $\text{SNR}_\text{thr}=10$. The gray dotted line is the joint accessibility SNR curve for $\text{SNR}_\text{thr}=10$. The red dotted line is the joint SNR curve for $\text{SNR}_\text{thr}=2$. Finally, the joint likelihood analysis curve is also shown for comparison to $\text{SNR}_\text{thr}=2$. Blue diamonds denote model choices whose spectra are shown in Figure \ref{Omega-curve}. The circles show benchmark points from various PT scenarios, taken from \cite{Caprini:2016re}.}
\label{contours}
\end{figure*}

\begin{figure*}[tp!]
\centering
\begin{subfigure}[t]{0.4\textwidth}
\includegraphics[width=\textwidth]{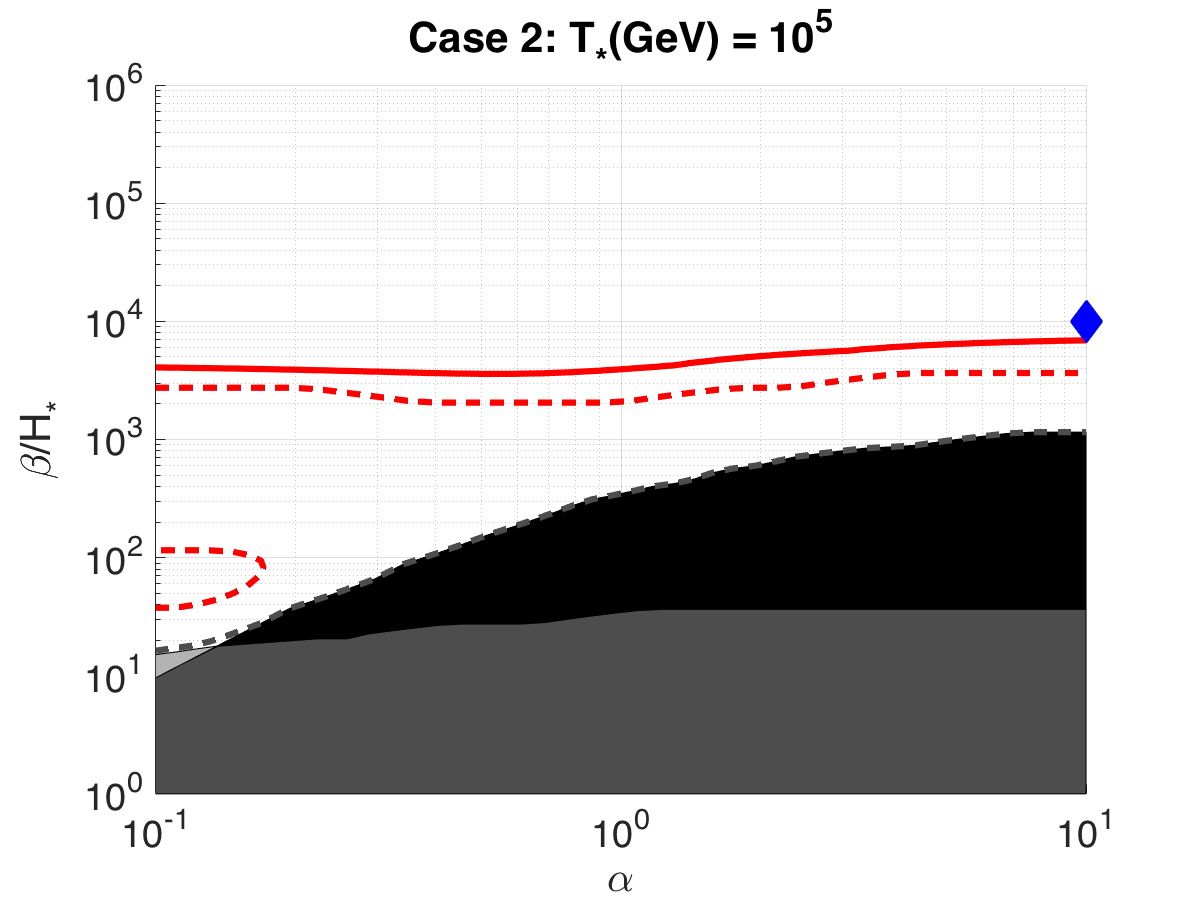}
\end{subfigure}
\begin{subfigure}[t]{0.4\textwidth}
\includegraphics[width=\textwidth]{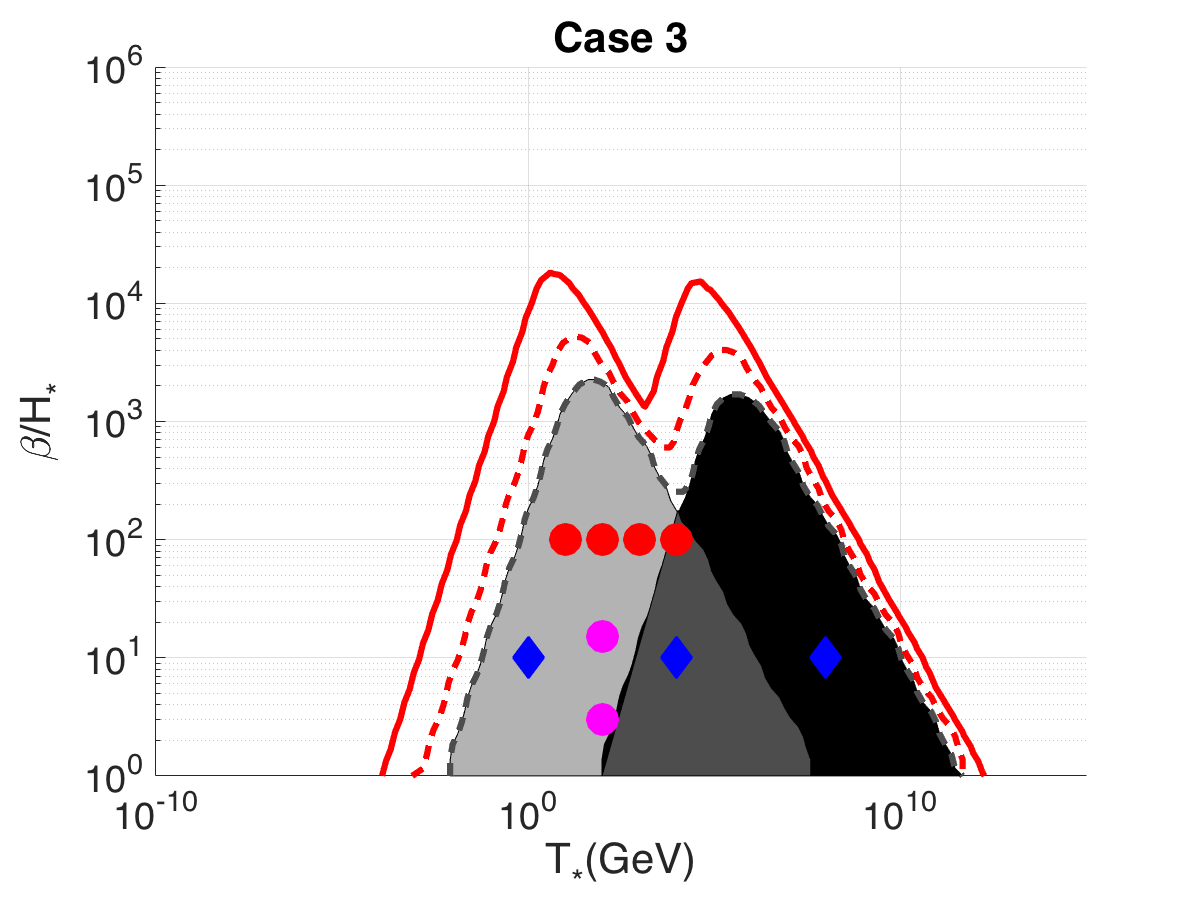}
\end{subfigure}
\caption{Shown here are contour plots for Cases 2 and 3. Note that Case 2, in which the scalar field plays a role, is plotted only for $\alpha > \alpha_{\infty}$. Similar notation for the curves and shaded regions is used as in Figure \ref{contours}.}
\label{contours2}
\end{figure*}

Results of our analysis can be seen in Figures \ref{contours} and~\ref{contours2}. Figure \ref{contours} shows the analysis of five temperature values for Case 1, while Figure \ref{contours2} shows the analysis done for cases 2 and 3. The shaded regions correspond to the SNR analysis done for $\text{SNR}_\text{thr}=10$. The light gray shaded regions are regions accessible to LISA, the black shaded regions are accessible to Cosmic Explorer, and the dark gray shaded regions are accessible to both operating separately. The gray dotted line shows the accessible region for both detectors operating simultaneously. The red dotted line shows the joint accessibility computed for $\text{SNR}_\text{thr}=2$, and the solid red line denotes the joint accessibility likelihood curve. 

Having both detectors operating allows wide access to many different regions of the parameter space that would be out of range of just one. For Cases 1 and 2, LISA is more sensitive to lower values of the parameter $\beta/H_\star$ while Cosmic Explorer is sensitive to higher values, allowing much of the chosen range of $\beta/H_\star$ to be accessible at higher values in the chosen range of $\alpha$. Additionally, the contours for Cases 1 and 3 show that while LISA is more sensitive to the background at lower temperatures, Cosmic Explorer allows better sensitivity at higher temperatures. These results can be understood by the fact that in general, higher temperature and $\beta/H_\star$ values shift the gravitational wave spectrum to higher frequencies, being stronger in the regions accessible to next-generation terrestrial detectors rather than LISA. 

Having regions of the parameter space accessible to both CE and LISA provides additional advantages in regards to constraining the spectrum. A joint detection in multiple frequency bands would be a major discovery, allowing better estimates of model parameters than would be possible with each detector separately. Similarly, a discovery by only one of the detectors would rule out corresponding areas of the parameter space.

It is evident that the method used for analysis has an impact on the computed detectable region, and therefore it is relevant to consider both. The likelihood analysis computes the greatest accessible region, while the SNR curves compute a smaller accessible region of the parameter space. However, the value of $\text{SNR}_\text{thr}$ chosen also has an impact; for $\text{SNR}_\text{thr}=2$ the curve is closer to the likelihood curve then for $\text{SNR}_\text{thr}=10$. 

The 95\% confidence likelihood curves should correspond to the analysis done for $\text{SNR}_{\text{thr}}=2$, because both are calculating $2\sigma$ confidence regions.  However, it is evident from the plots that the curves are not the same. For the likelihood analysis, probabilities are calculated on levels of a normalized likelihood curve and contours are made at the level $Z$ for which $95\%$ of the summed likelihood is above $Z$. These levels do not necessarily correspond to a constant SNR along the contours because the prior on these values is flat in the parameters $\alpha$, $\beta/H_\star$ and $T_\star$ rather than in $\Omega_{\text{GW}}$. For the SNR calculation, each individual point in the parameter space is considered for whether it exceeds the threshold SNR value, so it is ultimately a different calculation that yields a slightly different result. 
The shaded regions, corresponding to ${\rm SNR=10}$, identify parts of the parameter space where a detection with strong significance could be made based on the criteria outlined in this paper. The $95\%$ contours based on the likelihood model show where we would expect to be able to place $95\%$ upper limits in the absence of a detection. Backgrounds in between the $95\%$ upper limit and a strong detection would lead to a marginally detected signal; by integrating for a longer period of time, the confidence in signals would increase.

Finally, it is worth considering what these plots can tell us about detection possibilities for the gravitational wave signal from specific phase transition models. Extensions of the standard model that predict first order phase transitions at the electroweak scale are widely studied in the literature, some examples are supersymmetry~(see e.g.~\cite{Huber:2016re}), Higgs doublet models~(see e.g.~\cite{Dorsch:2016re}), and higher dimensional operators~(see e.g.~\cite{Huber:2008re}). Other scenarios beyond the electroweak scale that predict phase transitions include an additional boson field \cite{randall,konstandin} or a Dark Matter scenario \cite{schwaller}. 

The work done in \cite{Caprini:2016re} analyzed the detectability of specific benchmark points in the parameter space illustrating models considered in other studies. We include some of these points in Figures \ref{contours} and \ref{contours2} as well, shown by the colored circles in the contour plots. For Case 1, the temperature of the points is approximated to place it in a specific contour plot. As found in \cite{Caprini:2016re}, we confirm here that many of these points are in the accessible region of LISA, which may then provide valuable insight into new physics. 

In summary, we have seen how having both LISA and Cosmic Explorer together allows for wide detection possibilities for the gravitational wave background arising from phase transitions, as these detectors can probe complementary regions of the parameter space. We have used two different analysis methods to assess the detectability and stress that neither one necessarily provides the ''correct answer'' but that both should be considered and compared. The regions of the parameter space accessible to LISA and Cosmic Explorer include many predicted early universe models, opening up unique possibilities to study the early universe with gravitational wave observations. 

{\it Acknowledgements:} The work of MFA, AM, and VM was in part supported by the NSF grant
PHY1505870.

\bibliographystyle{apsrev}
\bibliography{references}

\end{document}